\pdfoutput=1

\documentclass[twocolumn]{aastex62}
\received{2018 January 30}
\revised{2018 April 04}
\accepted{2018 April 07}
\submitjournal{\apjl}

\shorttitle{H$_2$ Dissociation\slash Recombination in UHJs}
\shortauthors{Bell \& Cowan}

\usepackage{graphicx}
\usepackage{amsmath}
\usepackage{amssymb}
\usepackage{bm}
\usepackage{cleveref}
\newcommand{\RNum}[1]{\uppercase\expandafter{\romannumeral #1\relax}}
\DeclareRobustCommand{\rchi}{{\mathpalette\irchi\relax}}
\newcommand{\irchi}[2]{\raisebox{\depth}{$#1\chi$}} 

\newcommand*{\GtrSim}{\smallrel\gtrsim}
\newcommand*{\LessSim}{\smallrel\lesssim}

\makeatletter
\newcommand*{\smallrel}[2][.8]{%
  \mathrel{\mathpalette{\smallrel@{#1}}{#2}}%
}
\newcommand*{\smallrel@}[3]{%
  \sbox0{$#2\vcenter{}$}%
  \dimen@=\ht0 %
  \raise\dimen@\hbox{%
    \scalebox{#1}{%
      \raise-\dimen@\hbox{$#2#3\m@th$}%
    }%
  }%
}
\makeatother

\begin{document}

\title{Increased Heat Transport in Ultra-Hot Jupiter Atmospheres Through \\ H$_2$ Dissociation\slash Recombination}

\correspondingauthor{Taylor J. Bell}
\email{taylor.bell@mail.mcgill.ca}

\author[0000-0003-4177-2149]{Taylor J. Bell}
\altaffiliation{McGill Space Institute; Institute for Research on Exoplanets;\\ Centre for Research in Astrophysics of Quebec}
\affil{Department of Physics, McGill University, 3600 rue University, Montr\'eal, QC H3A 2T8, Canada}

\author[0000-0001-6129-5699]{Nicolas B. Cowan}
\altaffiliation{McGill Space Institute; Institute for Research on Exoplanets;\\ Centre for Research in Astrophysics of Quebec}
\affil{Department of Physics, McGill University, 3600 rue University, Montr\'eal, QC H3A 2T8, Canada}
\affil{Department of Earth \& Planetary Sciences, McGill University, 3450 rue University, Montr\'eal, QC H3A 0E8, Canada}




\begin{abstract}
	A new class of exoplanets is beginning to emerge: planets whose dayside atmospheres more closely resemble stellar atmospheres as most of their molecular constituents dissociate. The effects of the dissociation of these species will be varied and must be carefully accounted for. Here we take the first steps towards understanding the consequences of dissociation and recombination of molecular hydrogen (H$_2$) on atmospheric heat recirculation. Using a simple energy balance model with eastward winds, we demonstrate that H$_2$ dissociation\slash recombination can significantly increase the day--night heat transport on ultra-hot Jupiters (UHJs): gas giant exoplanets where significant H$_2$ dissociation occurs. The atomic hydrogen from the highly irradiated daysides of UHJs will transport some of the energy deposited on the dayside towards the nightside of the planet where the H atoms recombine into H$_2$; this mechanism bears similarities to latent heat. Given a fixed wind speed, this will act to increase the heat recirculation efficiency; alternatively, a measured heat recirculation efficiency will require slower wind speeds after accounting for H$_2$ dissociation\slash recombination.
\end{abstract}



\section{Introduction}
Most gas giant exoplanets have atmospheres dominated by molecular hydrogen (H$_2$). However, on planets where the temperature is sufficiently high, a significant fraction of the H$_2$ will thermally dissociate; one may call these planets ultra-hot Jupiters (UHJs). Only a handful of known planets have dayside temperatures this high, but the TESS mission is expected to discover hundreds more as it includes many early-type stars \citetext{George Zhou, private communication 2017}. These UHJs are an interesting intermediate between stars and cooler planets, and they will allow for useful tests of atmospheric models.

At these star-like temperatures, the H$^-$ bound-free and free-free opacities should play an important role in the continuum atmospheric opacity which has recently been detected in dayside secondary eclipse spectra \citep{bell2017, arcangeli2018}. These recently reported detections of H$^-$ opacity provide evidence that H$_2$ is dissociating in the atmospheres of gas giants at this temperature range. However, the thermodynamical effects of H$_2$ dissociation\slash recombination have yet to be explored.

Both theoretically \citep[e.g.][]{perez-becker2013, komacek2016} and empirically \citep[e.g.][]{zhang2018, schwartz2017}, we expect the day--night temperature contrast on hot Jupiters to increase with increasing stellar irradiation; temperature gradients \mbox{$\GtrSim 1000$\,K} can be expected for UHJs. As temperatures vary drastically between day and night, the local thermal equilibrium (LTE) H$_2$ dissociation fraction will also vary. The recombination of H into H$_2$ is a remarkably exothermic process, releasing \mbox{$q=$\ 2.14$\times$10$^8$\,J\,kg$^{-1}$} \citep{dean1999}; this is 100$\times$ more potent than the latent heat of condensation for water. For reference, latent heat is responsible for approximately half of the heat recirculation on Earth ($L/(c_p \Delta T) \sim 1$), while the effect of H$_2$ dissociation\slash recombination should be even stronger for UHJs ($q/(c_p \Delta T) \sim 10^2$).

\begin{figure}
	\centering
	``Top-Down'' View\\
	\vspace{0.25cm}
	\includegraphics[width=0.7\linewidth]{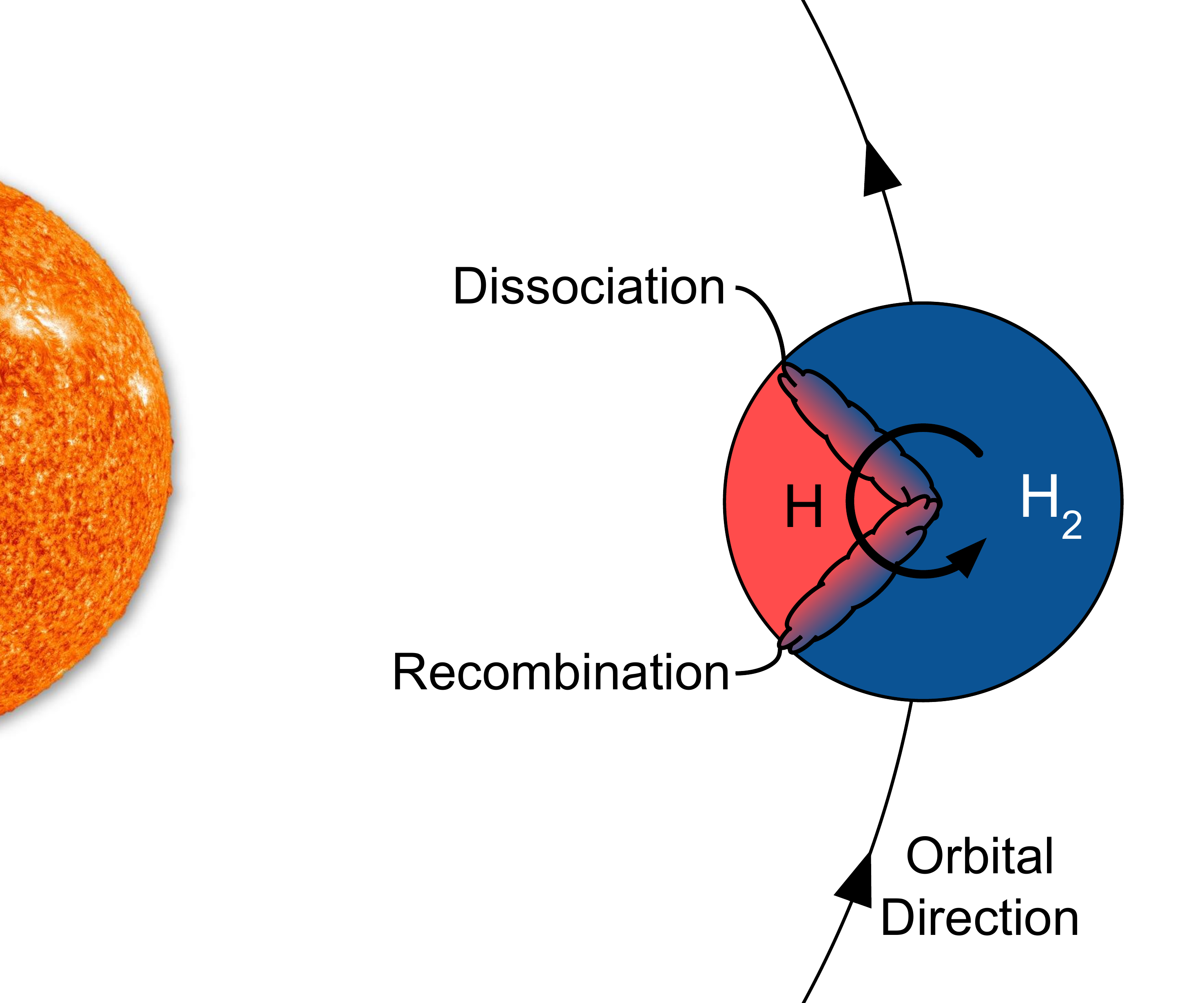}
	\caption{A cartoon showing a ``top-down'' view of the expected dissociation and recombination of H$_2$ on an ultra-hot Jupiter (UHJ). The orbital direction and the direction of winds on the planet are indicated with black arrows.} \label{fig:cartoon}
\end{figure}

Building on this intuition, we might expect that H will recombine into H$_2$ as gas carried by winds flows eastward from the sub-stellar point, significantly heating the eastern hemisphere of the planet. As the gas continues to flow around to the dayside, the H$_2$ will again dissociate and significantly cool the western hemisphere. A cartoon depicting this layout is shown in \Cref{fig:cartoon}. If unaccounted for while modelling a phasecurve, this may manifest itself as an ``unphysically'' large eastward offset as was previously reported for WASP-12b \citep{cowan2012}.

A large number of circulation models have been developed for studying exoplanet atmospheres, ranging from simple energy balance models \citep[e.g.][]{cowan2011b} to more advanced general circulation models \citep[e.g.][]{showman2009, rauscher2010, amundsen2014, zhangX2017, heng2017, dobbs-dixon2017}. To our knowledge, however, no published general circulation models account for the cooling\slash heating due to the energies of H$_2$ dissociation\slash recombination \citep[although some planet formation models do account for this, e.g.][]{berardo2017a}. Here we aim to qualitatively explore the effects of H$_2$ dissociation\slash recombination using a simple energy balance model adapted from that described by \cite{cowan2011b}, using code based on that implemented by \citet{schwartz2017}. We leave it to those with more advanced circulation models to explore this problem in a more rigorous and quantitative manner.

\section{Energy Transport Model}
\subsection{Heating Terms}
First, let $\epsilon$ be the energy per unit area of a parcel of gas. Ignoring H$_2$ dissociation/recombination and any internal heat sources, and assuming the gas parcel cools radiatively, energy conservation gives
\begin{equation*}
	\frac{d\epsilon}{dt} = F_{\rm in} - F_{\rm out},
\end{equation*}
with $F_{\rm{in}}$ and $F_{\rm{out}}$ given by
\begin{equation*}
	F_{\rm{in}} = (1-A_B)F_*\sin{\theta}\max{\Big(\cos{\Phi(t)}, 0\Big)},
\end{equation*}
\begin{equation*}
	F_{\rm{out}} = \sigma T^4.
\end{equation*}
The planet's Bond albedo is given by $A_B$, $\theta$ is the co-latitude of the gas parcel, $T$ is the temperature of the gas parcel, and $\sigma$ is the Stefan-Boltzmann constant. The incoming stellar flux is given by \mbox{$F_* = \sigma T_{*, \rm{eff}}^4 (R_*/a)^2$}, where $T_{*, \rm{eff}}$ is the stellar effective temperature, $R_*$ is the stellar radius, and $a$ is the planet's semi-major axis. The stellar hour angle, $\Phi(t)$, incorporates both advection and planetary rotation.

In order to include H$_2$ dissociation/recombination, we add a new term accounting for the energy flux from these effects. This can be done with
\begin{equation}
	\frac{d\epsilon}{dt} = F_{\rm in} - F_{\rm out} - \frac{d\mathbb{Q}}{dt},
	\label{eqn:e1}
\end{equation}
where the energy per unit area stored by H$_2$ dissociation is given by
\begin{equation*}
	\mathbb{Q} = q \rchi_{\rm H} \Sigma,
\end{equation*}
where $\Sigma$ is the mass per unit area of H and H$_2$ in the parcel of gas (in kg\,m$^{-2}$), $q=$~2.14$\times$10$^8$\,J\,kg$^{-1}$ is the H$_2$ bond dissociation energy per unit mass at 0\,K, and $\rchi_{\rm H}$ is the dissociation fraction of the gas. \mbox{$\rchi_{\rm H}=1$} means the gas is completely dissociated (all atomic). Assuming the gas parcel is in hydrostatic equilibrium, we can use
\begin{equation*}
	\Sigma = \int_{z_0}^{\infty} \rho(z) dz = (P_0/g)
\end{equation*}
where $z_0$ is some reference height, $P_0$ is the atmospheric pressure corresponding to $z_0$, and $\rho$ is the density of the gas. This then allows us to rewrite $\mathbb{Q}$ as
\begin{equation*}
	\mathbb{Q} = (P_0/g) q \rchi_{\rm H}.
\end{equation*}

The time derivative of $\mathbb{Q}$ is then
\begin{equation}
	\frac{d\mathbb{Q}}{dt} = (P_0/g) q \frac{d\rchi_{\rm H}}{dt} = (P_0/g) q \frac{d\rchi_{\rm H}}{dT}\biggr\rvert_T \frac{dT}{dt},
	\label{eqn:e2}
\end{equation}
where we have assumed the gas parcel's $P_0/g$ remains constant, and where we have made use of the chain rule to expand $d\rchi_{\rm H}/dt$.%

We model the LTE H$_2$ dissociation fraction by solving the Saha equation as stated in Appendix A of \cite{berardo2017a} for $\rchi_{\rm H}$, assuming the atmosphere consists of only H and H$_2$:
\begin{equation}
	\rchi_{\rm H}(P,T) = \frac{n_{\rm H}}{n_{\rm H}+n_{\rm H_2}} = \frac{2}{1+\sqrt{1+4Y}},
	\label{eqn:chi}
\end{equation}
where $n_{\rm H}$ and $n_{\rm H_2}$ are the number densities of H and H$_2$,
\begin{equation*}
	Y = \frac{T^{-3/2} \exp(q/2 m_{\rm H} k_{\rm B}T) P}{2(\pi m_{\rm H})^{3/2} k_{\rm B}^{5/2} h^{-3} \Theta_{\rm rot}},
	\label{eqn:y}
\end{equation*}
where $m_{\rm H}$ is the mass of the hydrogen atom, $k_{\rm B}$ is the Boltzmann constant, $h$ is the Planck constant, \mbox{$\Theta_{\rm rot}=85.4$\,K} is rotational temperature of H$_2$ \citep{hill1986}, $P$ is the gas pressure, and $T$ is the temperature of the gas (in K). The LTE dissociation fraction is plotted in the top panel of \Cref{fig:cpVSh2}.

We can then find $d\rchi_{\rm H}/dT$ using the chain rule:
\begin{equation*}
	\frac{d\rchi_{\rm H}}{dT} = \frac{d\rchi_{\rm H}}{dY}\frac{dY}{dT}.
\end{equation*}
After some simplification, we then determine
\begin{equation}
	\frac{d\rchi_{\rm H}}{dT} = \frac{\displaystyle \rchi_{\rm H}^2 Y \bigg(\frac{3}{2}T^{-1} + (q/2 m_{\rm H} k_{\rm B}) T^{-2}\bigg)}{\displaystyle \sqrt{1+4Y}}.
	\label{eqn:dchi}
\end{equation}
To a good degree of accuracy, \Cref{eqn:chi,eqn:dchi} can be approximated at $P=0.1$\,bar using
\begin{equation}
	\rchi_{\rm H}(0.1\,\text{bar},T) = \frac{1}{2}\Bigg(1 + \text{erf}\bigg(\frac{T-\mu}{\sigma\sqrt{2}}\bigg)\Bigg)
\end{equation}
and
\begin{equation}
	\frac{d\rchi_{\rm H}(0.1\,\text{bar}, T)}{dT} = \frac{1}{\sigma\sqrt{2\pi}} e^{-(T-\mu)^2 \left/ 2\sigma^2 \right.}
\end{equation}
where $\sigma = 471$\,K and $\mu = 3318$\,K, and $\text{erf}$ is the error function; this approximation offers a 70\% increase in computation speed. It should be noted that we assume that this H$_2$ dissociation\slash recombination occurs instantaneously since the timescale in the temperature regime of UHJs at 0.1\,bar is $\sim$10$^{-3}$\,s \citep{rink1962, shui1973}.

\subsection{Thermal Energy}
We assume that the planet's energy is stored entirely as thermal energy
, as is done in other simple energy balance models \citep[e.g.][]{cowan2011b, pierrehumbert2010}. This assumption means
\begin{align}
	\frac{d\epsilon}{dt} &= \frac{d}{dt}(c_p T \Sigma) = \frac{d}{dt}\Big((P_0/g)c_p T\Big) \nonumber\\
	&= (P_0/g) \bigg( c_p \frac{dT}{dt} + T \frac{d c_p}{dt} \bigg) \nonumber\\
	&= (P_0/g) \frac{dT}{dt} \bigg( c_p + T \frac{d c_p}{dT}\biggr\rvert_T \bigg).
	\label{eqn:e3}	
\end{align}
where $c_p$ is the specific heat capacity of the gas, we have again assumed the gas parcel's $P_0/g$ remains constant, and we have used the chain rule to expand $d c_p/dt$.

The specific heat capacity of the gas will change as a function of temperature due to the slightly different values for H and H$_2$ as well as the variations in the specific heat capacity of H$_2$ as a function of temperature \citep{chase1998}; any model properly accounting for H$_2$ dissociation should account for this effect. In our model, we assume the atmosphere is made entirely of hydrogen and model the specific heat capacity of the gas by assuming it is well mixed so that
\begin{equation*}
	c_p = c_{p, \text{H}}\rchi_{\rm H} + c_{p, \text{H}_2}(1-\rchi_{\rm H}),
\end{equation*}
where both $\rchi_{\rm H}$ and $c_{p, \text{H}_2}$ are functions of temperature. The temperature derivative of $c_p$ is then given by
\begin{align*}
	\frac{dc_p}{dT}\biggr\rvert_T = (c_{p, \text{H}}- c_{p, \text{H}_2})\frac{d\rchi_{\rm H}}{dT}\biggr\rvert_T. 
\end{align*}

\subsection{Putting Everything Together}
\begin{figure}
	\centering
	\includegraphics[width=1\linewidth]{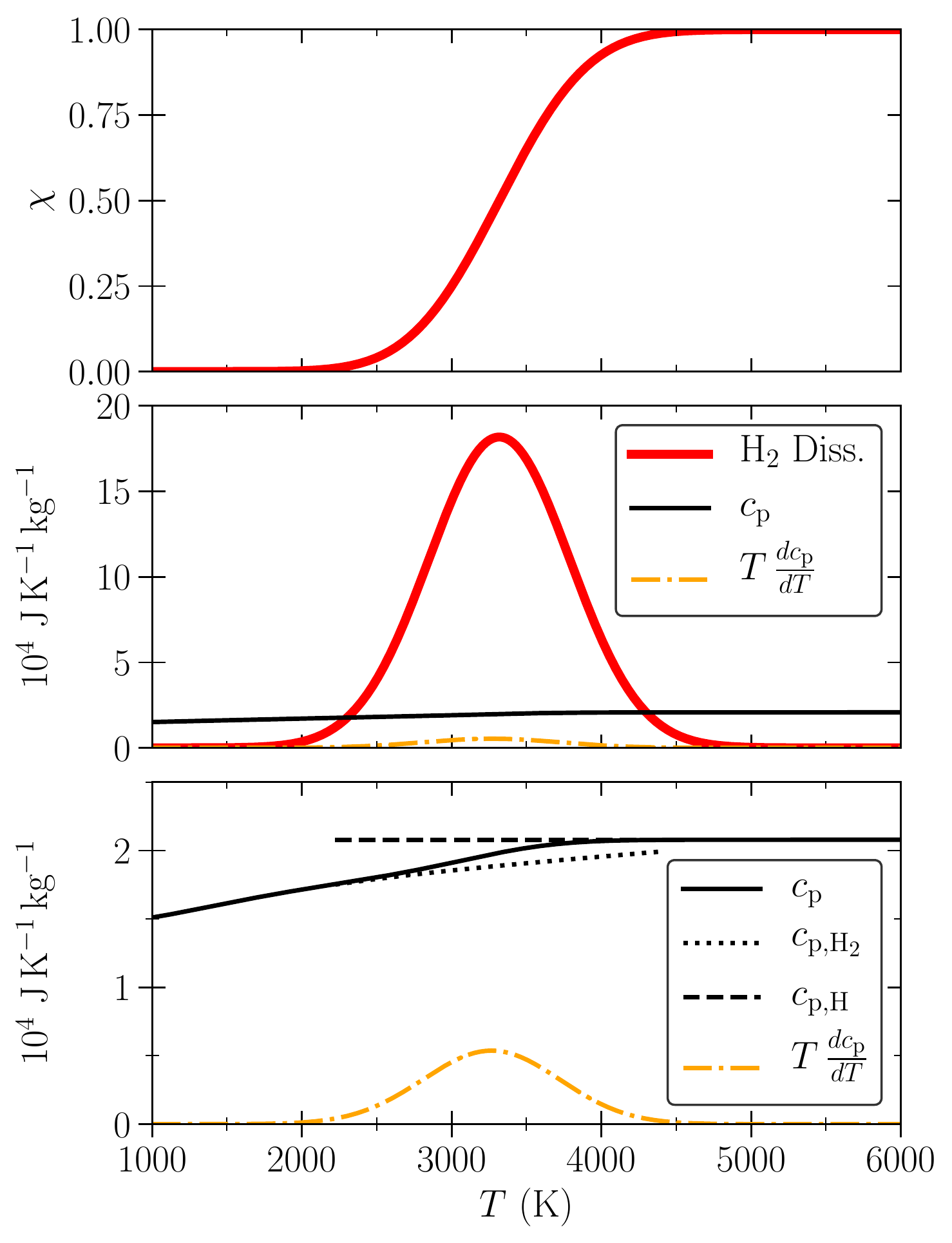}
	\caption{\textit{Top:} The LTE dissociation fraction of H$_2$ in a parcel of gas.
			 \textit{Middle:} A demonstration of the relative importance of the H$_2$ dissociation\slash recombination term in \Cref{eqn:dT}. For $2300 \LessSim T \LessSim 4300$, the energy absorbed by H$_2$ dissociation is greater than the energy stored as heat. Typical hot Jupiters are too cool to be affected by H$_2$ dissociation\slash recombination, but these processes should dominate on UHJs. \textit{Bottom:} An inset showing the specific heat capacity of a gas composed of H and H$_2$ in LTE (the same black line from the middle panel), the specific heat capacities of H and H$_2$ where they are able to exist in equilibrium, and the additional $T(dc_{\rm p}/dT)$ term. All panels assume a pressure of 0.1\,bar.}\label{fig:cpVSh2}
\end{figure}

Putting together \Cref{eqn:e1,eqn:e2,eqn:e3}, we get
\begin{align*}
	F_{\rm in} - F_{\rm out} - (P_0/g) \frac{dT}{dt} &\bigg(q \frac{d\rchi_{\rm H}}{dT}\biggr\rvert_T\bigg) \\
	= (P_0/g) \frac{dT}{dt} &\bigg(c_p + T \frac{d c_p}{dT}\biggr\rvert_T \bigg).
\end{align*}
After solving for $dT/dt$, we find
\begin{equation*}
	\frac{dT}{dt} = (F_{\rm in} - F_{\rm out}) (P_0/g)^{-1} \bigg(c_p + T \frac{d c_p}{dT}\biggr\rvert_T + q \frac{d\rchi_{\rm H}}{dT}\biggr\rvert_T\bigg)^{-1}.
\end{equation*}
Finally, a gas cell can then be updated using
\begin{equation}
	\Delta T = \frac{\Delta t(F_{\rm in} - F_{\rm out})}{\displaystyle (P_0/g)\bigg(c_p + T \frac{d c_p}{dT}\biggr\rvert_T + q \frac{d\rchi_{\rm H}}{dT}\biggr\rvert_T\bigg)}.
	\label{eqn:dT}
\end{equation}
Note that the entire sum in the denominator can instead be thought of as the specific heat capacity of a gas comprised of a mixture of H and H$_2$ in thermal equilibrium. The relative importance of the terms in this sum are shown in the bottom two panels of \Cref{fig:cpVSh2}.

\section{Simulated Observations and Qualitative Trends}

We now explore the effects of this new term in the differential equation governing the temperature of a gas cell. For this purpose, we create a latitude+longitude HEALPix grid where each parcel's temperature is updated using \Cref{eqn:dT} with code based on that developed by \citet{schwartz2017}.

While \cite{cowan2011b} were able to explore their model using dimensionless quantities, our updated model requires that we use dimensioned variables. We therefore adopt the values of the first discovered UHJ, \mbox{WASP-12b} \citep{hebb2009}. In particular, we set \mbox{$R_p = 1.90 \, R_J$,} \mbox{$a = 0.0234$\,AU,} \mbox{$M_p = 1.470 \, M_J$,} \mbox{$T_{*, \text{eff}} = 6360$\,K,} \mbox{$R_* = 1.657 \, R_\odot$,} \mbox{$P = 1.0914203$\,days} \citep{collins2017} and \mbox{$A_B = 0.27$} \citep{schwartz2017}. We have also assumed a photospheric pressure of 0.1\,bar, the approximate pressure probed by NIR observations of WASP-12b \citep{stevenson2014a}, which gives a radiative timescale of a few hours \citep[similar to the observed timescales for eccentric hot Jupiters, e.g.][]{lewis2013, deWit2016}. 
Wind speeds for \mbox{WASP-12b} have not been directly measured, but typical values for hot Jupiters are on the order of 1\,km\,s$^{-1}$ \citep[e.g.][]{koll2017}; for that reason, we focus on wind speeds around this order of magnitude.

\begin{figure*}[!htbp]
	\centering
	\begin{minipage}[c]{0.14\textwidth}
		\centering
		~
	\end{minipage}%
	\begin{minipage}[c]{0.25\textwidth}
		\centering
		~
	\end{minipage}%
	\begin{minipage}[c]{0.25\textwidth}
		\centering
		Equatorial Zonal Wind Speeds
	\end{minipage}%
	\begin{minipage}[c]{0.25\textwidth}
		\centering
		~
	\end{minipage}%
	\begin{minipage}[c]{0.08\textwidth}
		\centering
		~
	\end{minipage}
	
	\begin{minipage}[c]{0.14\textwidth}
		\centering
		~
	\end{minipage}%
	\begin{minipage}[c]{0.25\textwidth}
		\centering
		0.1~km\,s$^{-1}$
	\end{minipage}%
	\begin{minipage}[c]{0.25\textwidth}
		\centering
		1.0~km\,s$^{-1}$
	\end{minipage}%
	\begin{minipage}[c]{0.25\textwidth}
		\centering
		10~km\,s$^{-1}$
	\end{minipage}%
	\begin{minipage}[c]{0.08\textwidth}
		\centering
		~
	\end{minipage}%
	\vspace{0.1cm}
	
	\begin{minipage}[c]{0.14\textwidth}
		\centering
		Temperature Maps
	\end{minipage}%
	\begin{minipage}[c]{0.25\textwidth}
		\centering
		\includegraphics[width=\linewidth, trim=0 0 190 0, clip]{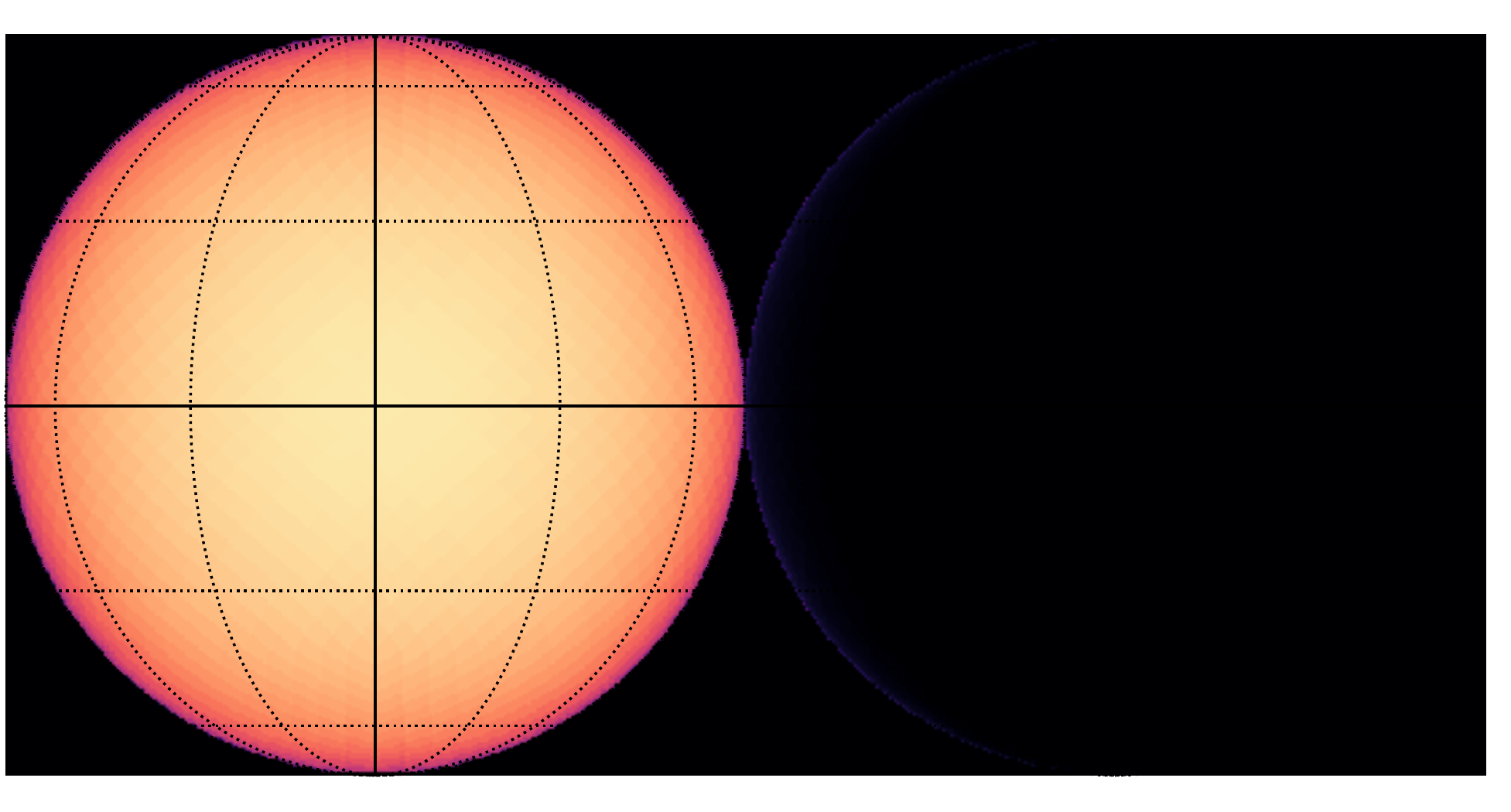}
	\end{minipage}%
	\begin{minipage}[c]{0.25\textwidth}
		\centering
		\includegraphics[width=\linewidth, trim=0 0 190 0, clip]{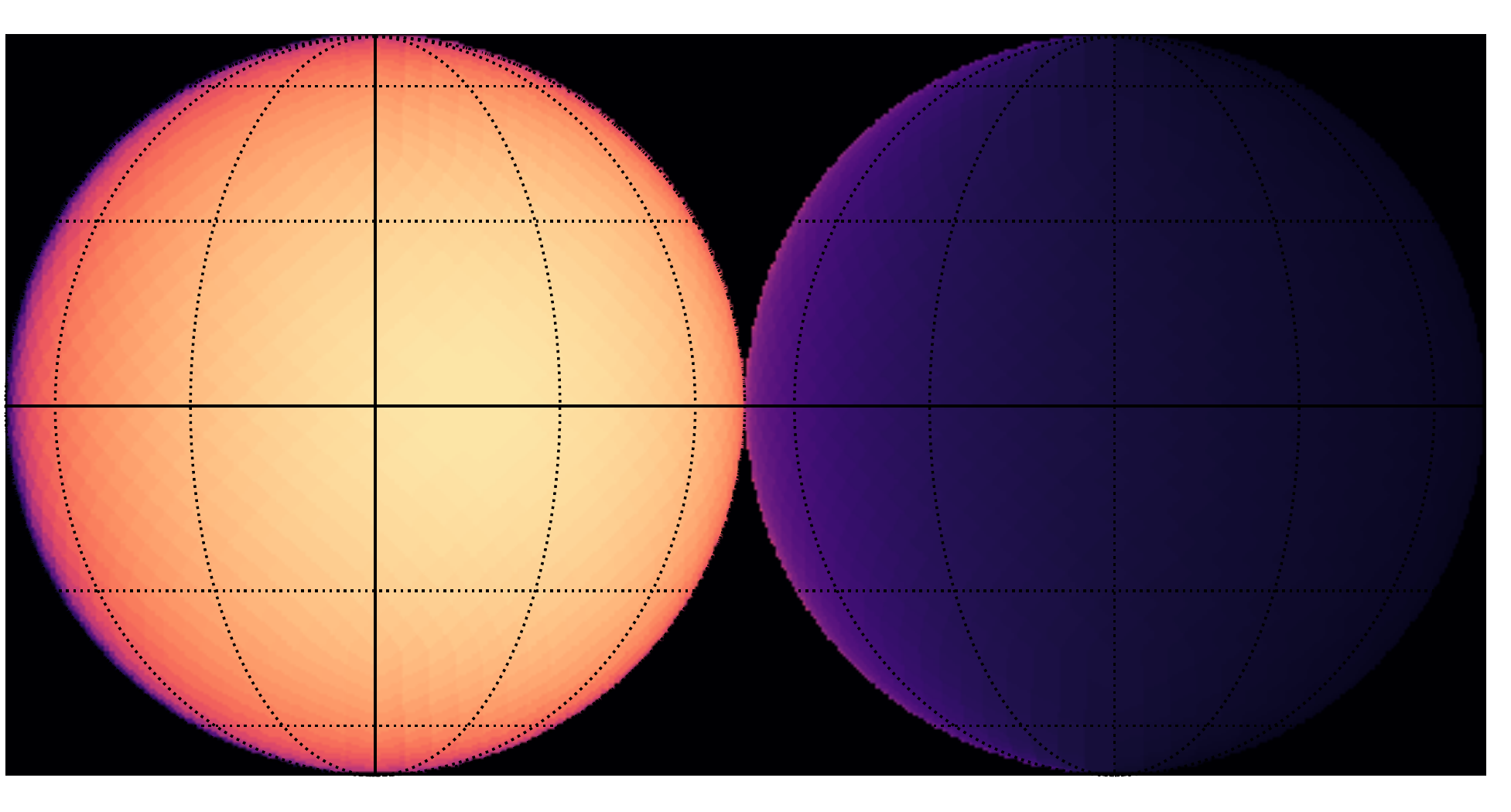}
	\end{minipage}%
	\begin{minipage}[c]{0.3345\textwidth}
		\centering
		\includegraphics[width=\linewidth, trim=0 0 0 0, clip]{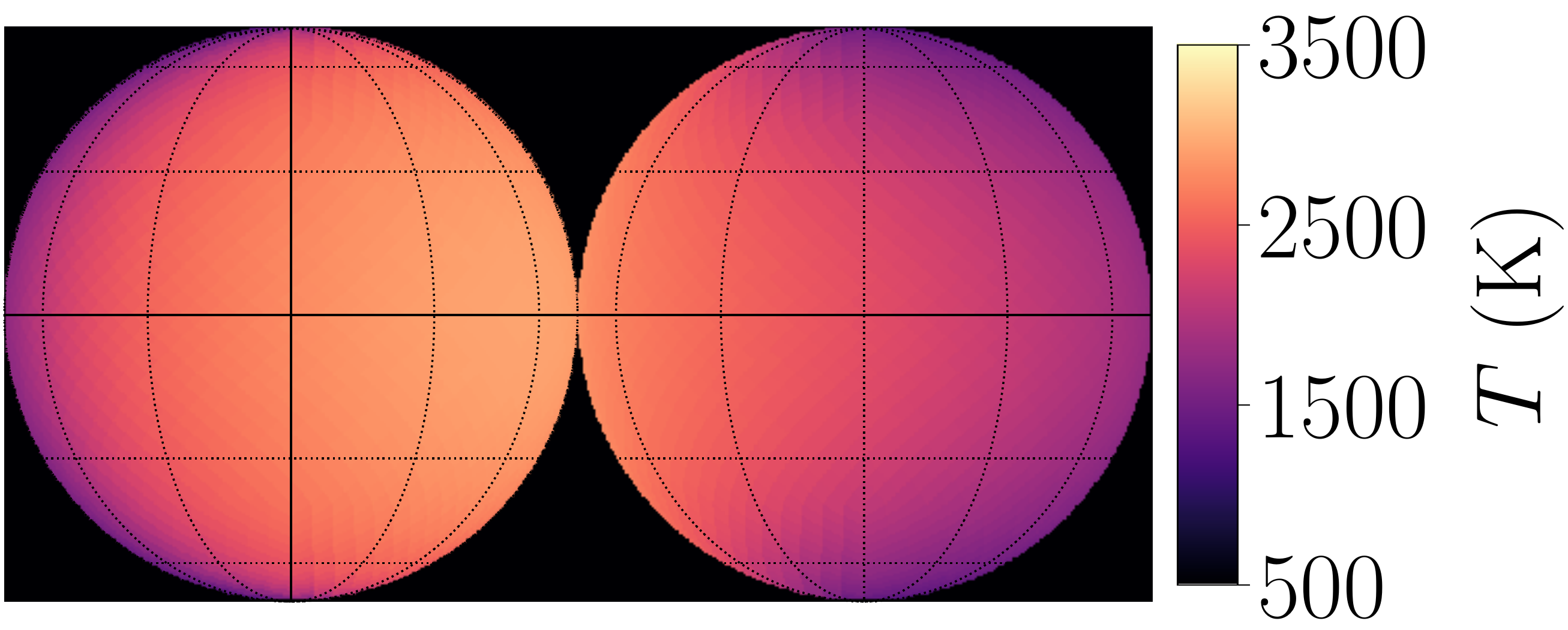}
	\end{minipage}%
	\vspace{0.1cm}
	
	\begin{minipage}[c]{0.14\linewidth}
		\centering
		H$_2$ Dissociation Maps
	\end{minipage}%
	\begin{minipage}[c]{0.25\linewidth}
		\centering
		\includegraphics[width=\linewidth, trim=0 0 190 0, clip]{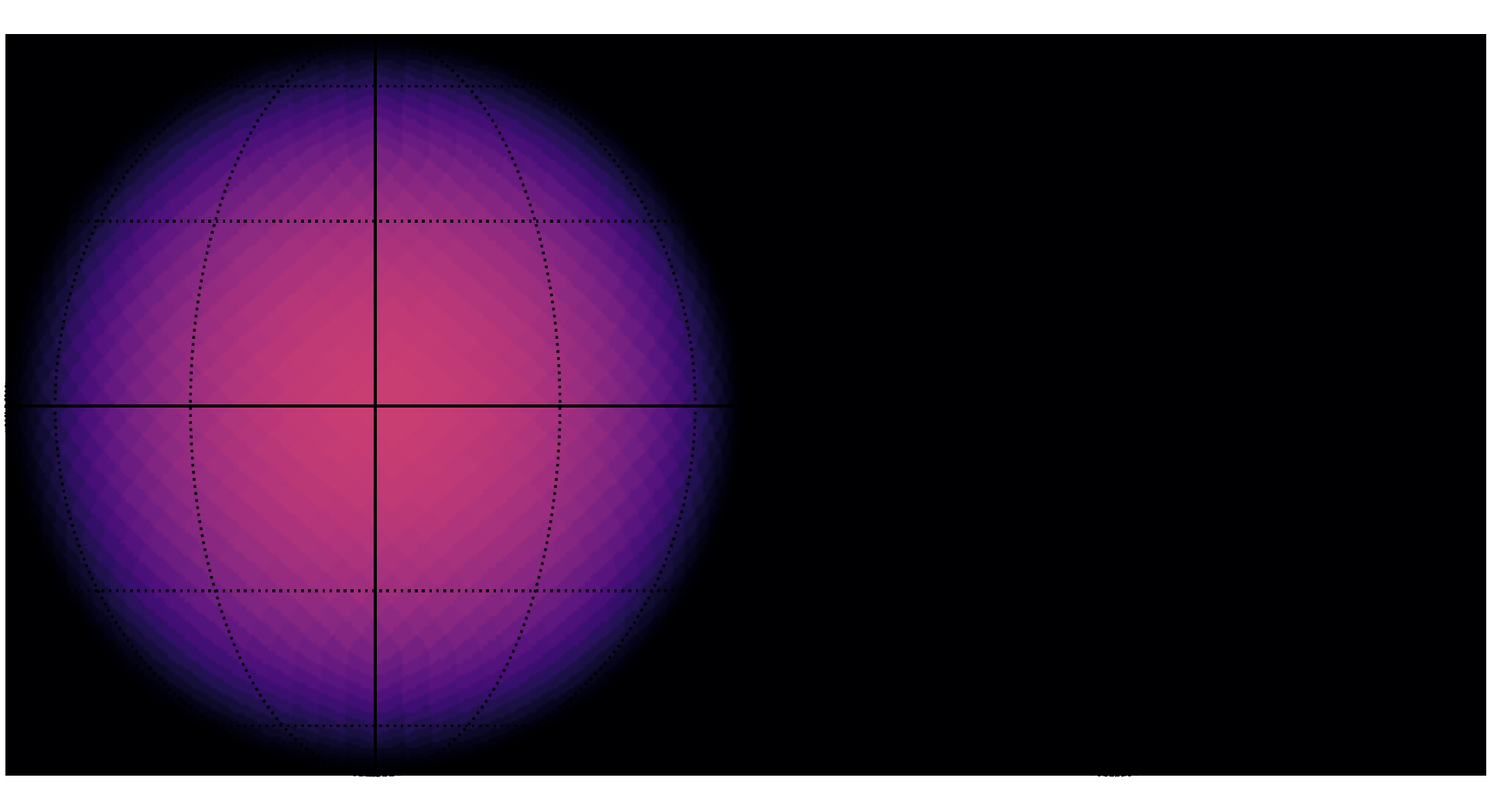}
	\end{minipage}%
	\begin{minipage}[c]{0.25\linewidth}
		\centering
		\includegraphics[width=\linewidth, trim=0 0 190 0, clip]{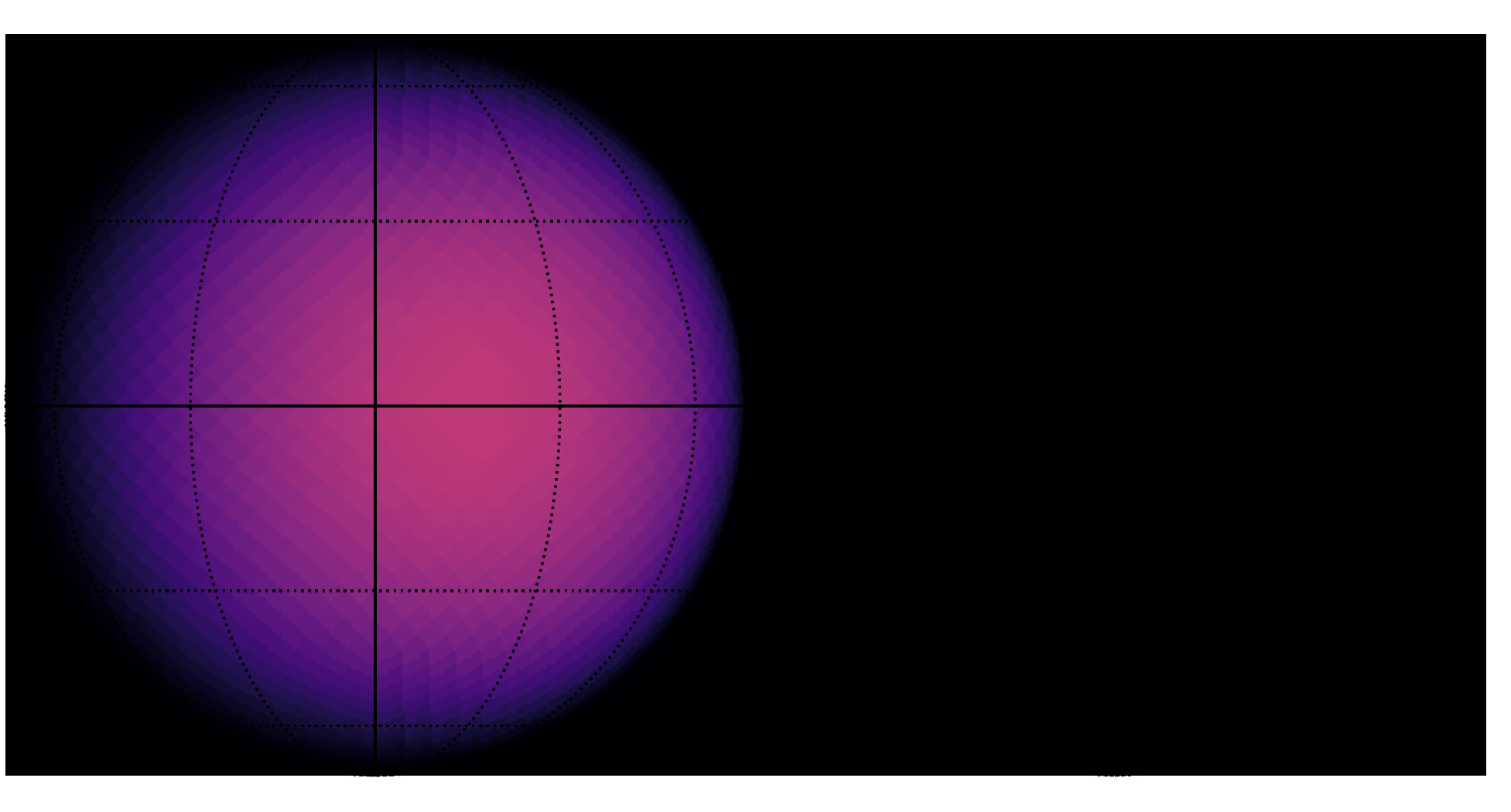}
	\end{minipage}%
	\begin{minipage}[c]{0.3345\linewidth}
		\centering
		\includegraphics[width=\linewidth, trim=0 0 0 0, clip]{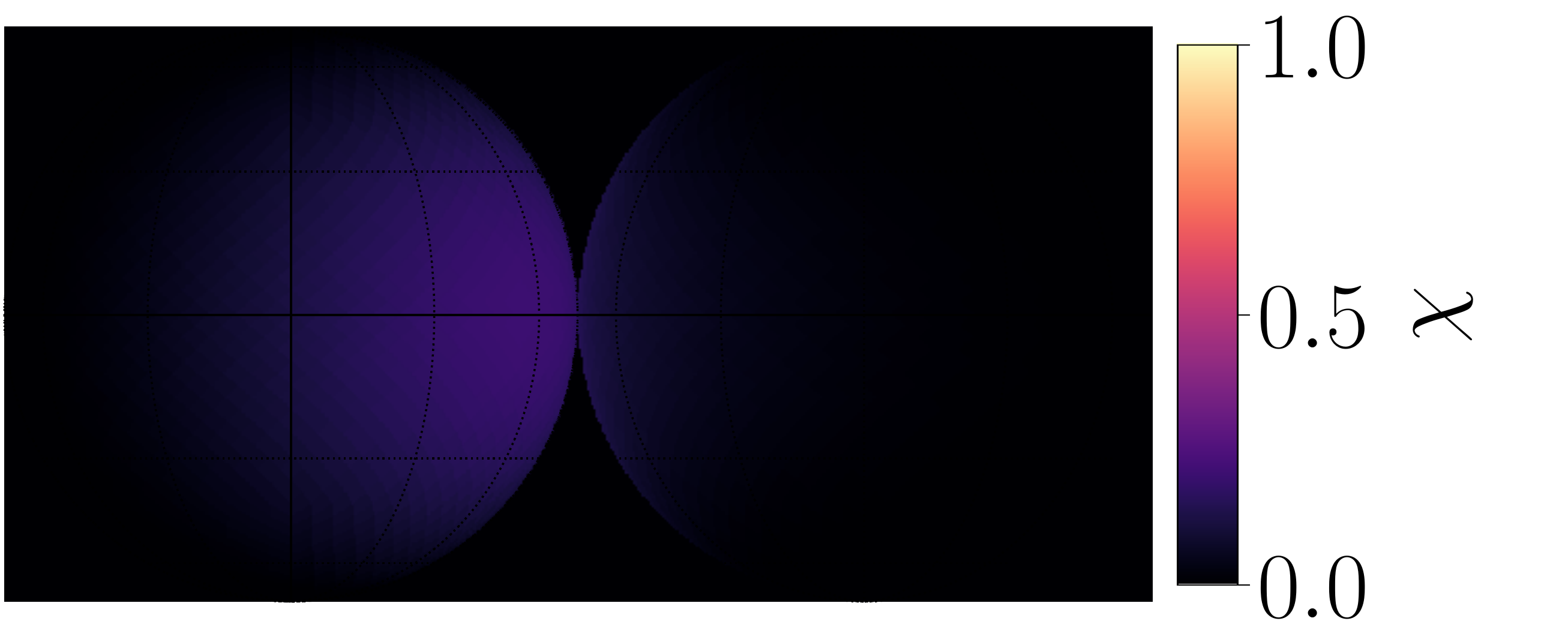}
	\end{minipage}
	\caption{Planetary maps, showing temperature and H$_2$ dissociation fraction, assuming different eastward zonal wind speeds. The dayside hemisphere is shown on the left side of each map with north at the top.}\label{fig:maps}

	\vspace{0.18cm}

	\begin{minipage}[c]{0.47\linewidth}
		\centering
		\includegraphics[width=\linewidth]{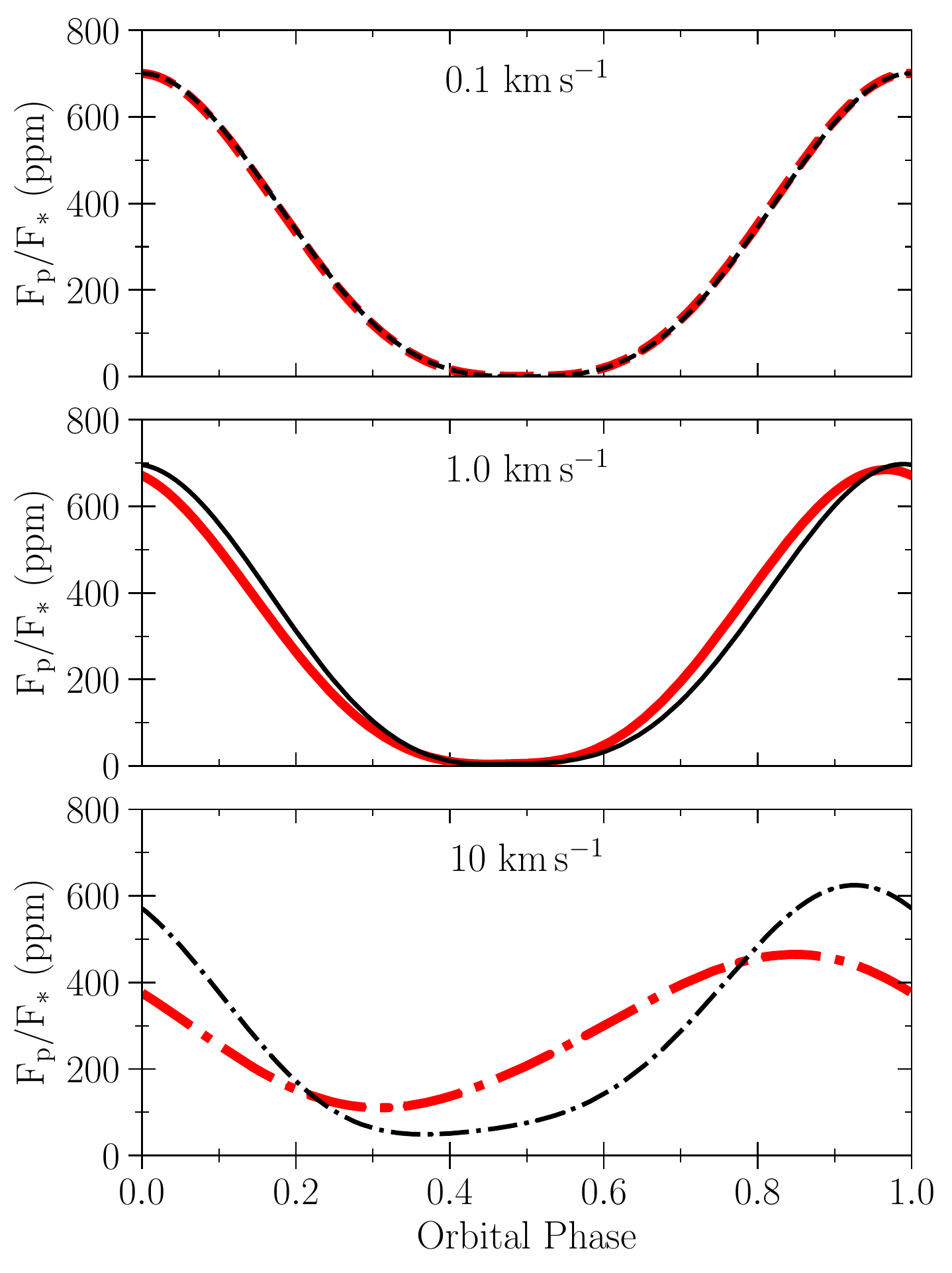}
		\caption{Model bolometric phasecurves assuming different eastward zonal wind velocities, ignoring eclipses and transits. The thick, red lines show the expected phasecurve accounting for H$_2$ dissociation\slash recombination, while the thinner, black models neglect these processes. Secondary eclipse would occur at a phase of 0.0, while a transit would occur at 0.5.}\label{fig:phasecurves}
	\end{minipage}\hfill%
	\begin{minipage}[c]{0.47\linewidth}
		\centering
		\includegraphics[width=\linewidth]{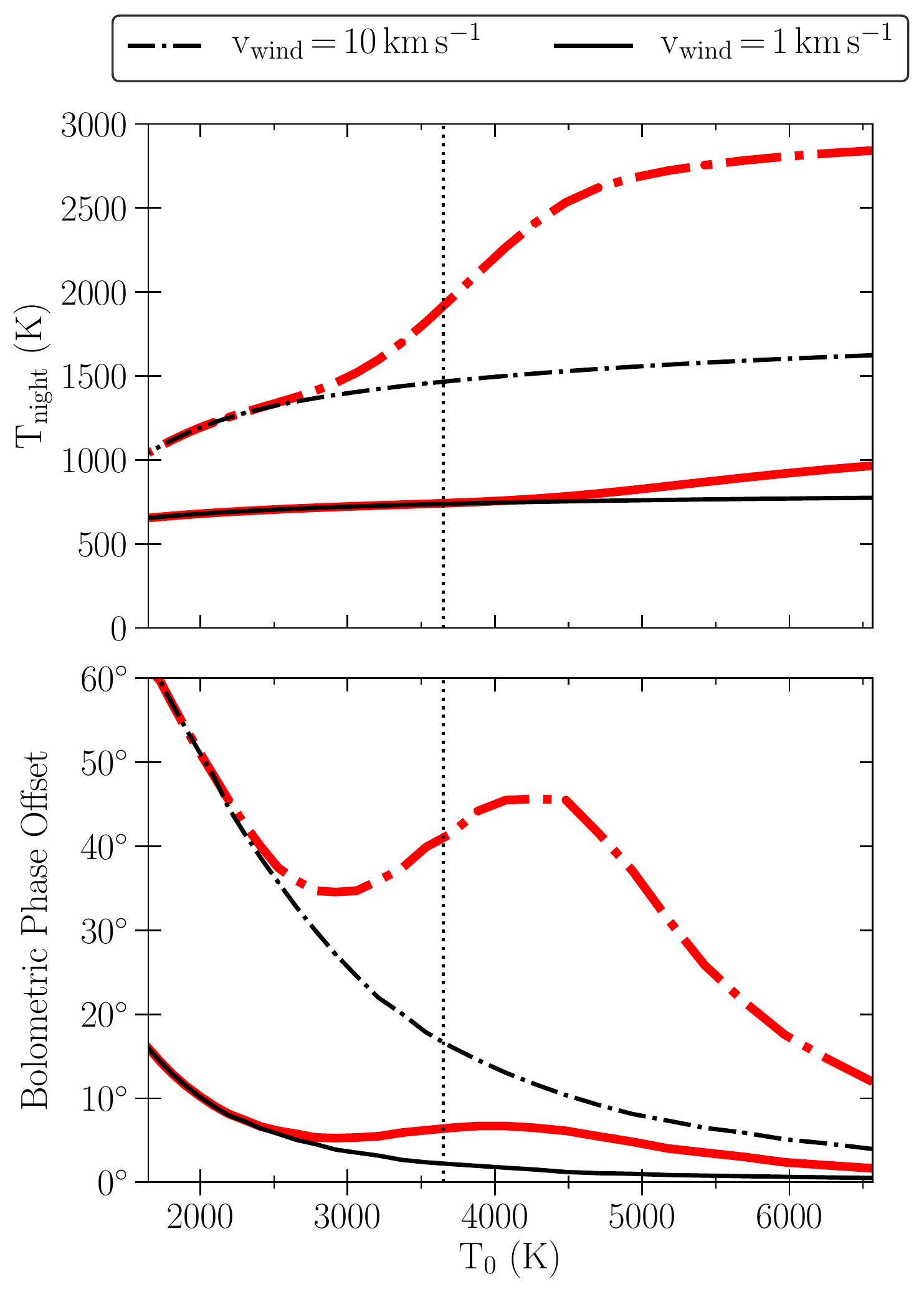}
	\caption{A figure showing the trends in nightside apparent temperature and phase offset as a function of irradiation temperature ($T_0 \equiv T_{\rm*,eff}\sqrt{R_{*}/a}$), given theoretical bolometric phasecurve measurements. Thick, red lines show models including H$_2$ dissociation\slash recombination for WASP-12b, while thin, black lines show models neglecting these effects. Models sharing the same wind speed share linestyles, and all models assume a Bond albedo of 0.3 \citep[which is typical for hot Jupiters;][]{zhang2018,schwartz2017}. A vertical dotted line shows the location of WASP-12b.}\label{fig:TcontVsOff}
	\end{minipage}
	
\end{figure*}

First, let's explore the effects of H$_2$ dissociation\slash recombination at a spatially resolved scale. \Cref{fig:maps} shows temperature and H$_2$ dissociation maps for three different wind speeds. In the limit of infinite wind speeds, there will be no temperature gradients and H$_2$ dissociation\slash recombination will not play a role. In the limit of an atmosphere in radiative equilibrium (wind speed = 0), there will be no variation in the temperature of a parcel and H$_2$ dissociation\slash recombination will play no role. Outside of these two unphysical limits, H$_2$ dissociation\slash recombination will always be occurring somewhere on UHJs.

We now consider phasecurve observations --- this requires that we convolve the planet map with a visibility kernel at each orbital phase \citep{cowan2013}, which acts as a low-pass filter. \Cref{fig:phasecurves} shows model phasecurves for three wind speed; this figure shows that H$_2$ dissociation\slash recombination can have a significant effect. At a constant wind speed, the first obviously affected observable when accounting for dissociation\slash recombination is the increased offset of the peak in the phasecurve towards the east (the same direction as the prescribed wind). Another affected observable is the amplitude of the phase variations which is reduced when H$_2$ dissociation\slash recombination is included. Also, a Fourier decomposition shows that nearly all of the power in the phasecurves accounting for H$_2$ dissociation\slash recombination is in the first and second order Fourier series terms ($1f_{\text{orb}}$ and $2f_{\text{orb}}$). Finally, \Cref{fig:TcontVsOff} shows the trends in phase offset and nightside temperature for two wind speeds, both accounting for and neglecting H$_2$ dissociation\slash recombination.

\section{Model Assumptions}

With simplistic models, many important effects are necessarily swept under the rug. Here we aim to lift up the rug and shine a light on our assumptions to aid future work. While many of these assumptions will change the quantitative effects of H$_2$ dissociation\slash recombination, we expect that the overall qualitative impact of increased heat recirculation will be robust to these assumptions.

One important piece of physics that we have ignored (beyond a simple assumption of a 0.1\,bar photosphere) is atmospheric opacity. As \citet{dobbs-dixon2017} demonstrated, variations in opacity sources as a function of longitude can change the depth of the photosphere by an order of magnitude or more. Changing the H$_2$ dissociation fraction will change the importance of H$^-$ as an opacity source, and other standard opacity sources (e.g. H$_2$O and CO) will also likely be important, especially towards the cooler nightside. The insignificant detection of H$_2$O on the dayside of WASP-12b \citep{stevenson2014a} but significant detection in the planet's transmission spectrum \citep{kreidberg2015} clearly demonstrates that opacity sources should be expected to change on UHJs. Several of the standard molecular opacity sources will also overlap with the far broader H$^-$ absorption, which complicates a definitive detection of H$^-$ using broadband photometry, such as with \textit{Spitzer}/IRAC. The formation of clouds on the nightside of the planet would further complicate the interpretation of observed phasecurves, increasing the albedo of the west terminator while also insulating the nightside. While we have accounted for variations in the radiative timescale as a function of temperature, we have not accounted for changes due to varying opacity sources.

Additionally, as we have assumed all photons are emitted at a 0.1\,bar photosphere, the effects of the atmosphere's \mbox{T-P} profile have been neglected. As the H$_2$ dissociation fraction has a fairly weak dependence on gas pressure, the bulk of vertical variations in the H$_2$ dissociation fraction will likely be controlled by the vertical temperature gradient. Due to the lower density of the dissociated gas, one may expect vertical advection on UHJs where temperature decreases with altitude. Interestingly, however, observations of most UHJs are best explained by atmospheres with thermal inversions \citep{arcangeli2018, evans2017} or at least approximately isothermal profiles on the dayside \citep{crossfield2012, cowan2012}. Any non-isothermal \mbox{T-P} profile will alter the specifics of how efficiently heat is redistributed across the planet as different layers in a gas column will dissociate\slash recombine at different locations. Also, as we have neglected atmospheric opacity, we have assumed that each gas parcel emits as a blackbody with a single temperature.

Further, due to the changing scale height of the atmosphere at different latitudes and longitudes due to changes in temperature and H$_2$ dissociation fraction, there will likely be a tendency for gas to flow away from the sub-stellar point both zonally and meridionally. This is not accounted for in our toy model and would require a general circulation model. Instead, we have chosen eastward winds as they are predicted, and seen, for most hot Jupiters \citep[e.g.][]{showman2002, zhang2018}, although there are some exceptions \citep[e.g.][]{dang2018}. Similarly, our assumption of solid-body atmospheric rotation is clearly an over simplification which will need to be addressed in future work. Our model is also unable to predict the wind speeds of UHJs which would require the implementation of various drag sources, such as magnetic drag which \citet{menou2012} suggests will dominate at these high temperatures.

Also, we have assumed all heating is due to H$_2$ dissociation and radiation from the host star, neglecting other heat sources such as residual heat from formation \citep[which should be negligible for planets older than 1 Gyr;][]{burrows2006} as well as tidal, viscous, and ohmic heating. We have also neglected the presence of helium which will partially dilute the strength of H$_2$ dissociation\slash recombination as only $\sim$80\% of the atmosphere will be hydrogen. Finally, we have assumed that the planet has a uniform albedo which will not be the case in general \citep[e.g.][]{esteves2013, demory2013, angerhausen2015, parmentier2016}.

\section{Discussion and Conclusions}

A new class of exoplanets is beginning to emerge: planets whose dayside atmospheres resemble stellar atmospheres as their molecular constituents thermally dissociate. The impacts of this dissociation will be varied and must be carefully accounted for. Here we have shown that the dynamical dissociation and recombination of H$_2$ will play an important role in the heat recirculation of ultra-hot Jupiters. In the atmospheres of ultra-hot Jupiters, significant H$_2$ dissociation occurs on the highly irradiated dayside, absorbing some of the incident stellar energy and transporting it towards the nightside of the planet where the gas recombines. Given a fixed wind speed, this will act to increase the heat recirculation efficiency; alternatively, a measured heat recirculation efficiency will require slower wind speeds once H$_2$ dissociation\slash recombination has been accounted for.

Both theoretically and observationally, it has been shown that increasing irradiation tends to lead to poorer heat recirculation \citep[e.g.][]{komacek2016, schwartz2017}. However, there are a few notable exceptions to this rule at high temperatures. Recently, \citet{zhang2018} reported a heat recirculation efficiency of $\varepsilon \sim 0.2$ for the UHJ WASP-33b which is far higher than would be predicted by theoretical and observational trends. WASP-12b may also possess an unusually high heat recirculation efficiency and exhibit a greater phase offset than would be expected from simple heat advection\footnote{Depending on the decorrelation method used to reduce the \textit{Spitzer}/IRAC data for WASP-12b, the planet either has $\varepsilon \sim 0$ or $\varepsilon \sim 0.5$ \citep{cowan2012, schwartz2017}; although the former is the preferred model, further observations are critical to definitively choose between these values and test the predictions made in this article.} \citep{cowan2012}. However, the power in the second order Fourier series terms from H$_2$ dissociation\slash recombination seems to make the phasecurve more sharply peaked and does not seem to be able to explain the double-peaked phasecurve seen for WASP-12b by \citet{cowan2012}. Also, while \citet{arcangeli2018} find evidence of H$_2$ dissociation\slash recombination in the atmosphere of WASP-18b, \citet{maxted2013} finds the planet has minimal day--night heat recirculation. Given the expected increase in heat recirculation due to H$_2$ dissociation\slash recombination, this suggests that WASP-18b has only moderate winds and/or is too cool for these processes to play a strong role in the heat recirculation of this planet. Finally, near-infrared observations of KELT-9b, the hottest UHJ currently known \citep{gaudi2017}, could provide a fantastic test of this theory in the very high temperature regime.

\acknowledgments

T.J.B.\ acknowledges support from the McGill Space Institute Graduate Fellowship and from the FRQNT through the Centre de recherche en astrophysique du Qu\'ebec. The atmospheric model we use in this work is based upon code originally developed by Diana Jovmir and Joel Schwartz. We also thank Gabriel Marleau and Ian Dobbs-Dixon for their helpful insights. We have also made use of free and open-source software provided by the Python, SciPy, and Matplotlib communities.

\listofchanges

\end{document}